\documentstyle[11pt,newpasp,twoside,epsf]{article}
\def\edcomment#1{\iffalse\marginpar{\raggedright\sl#1\/}\else\relax\fi}
\reversemarginpar

\begin{document}
\title{Search for Southern-Hemisphere Giant Radio Galaxies with SALT}

\author{J. Machalski, S. Zola, D. Kozie{\l}}
\affil{Obserwatorium Astronomiczne Uniwersytetu Jagiello{\'n}skiego, 
ul. Orla 171, 30-244 Krak{\'o}w, Poland }

\author{M. Jamrozy}
\affil{Radioastronomisches Institut der Universit\"{a}t Bonn, Auf 
dem H{\"u}gel 71, 53-121 Bonn, Germany}

\begin{abstract}
An extensive search for distant Giant radio galaxies 
on the southern hemisphere is justified. We emphasize the crucial role of
optical spectroscopy in determination of their basic physical parameters,
i.e. the distance, projected linear size, volume of their lobes or cocoon,
luminosity, etc., and argue that SALT will be the best instrument for such
a task.
\end{abstract}

\section{Introduction}
Radio galaxies (RGs) represent the largest single objects in the Universe.
Powered by an active galactic nucleus (AGN) jets emerge from the central
engine, which very likely is a super-massive black hole accreting matter
surrounding it. There is a huge range of linear extent of the RGs: from
less than $10^{2}$ pc -- Gigahertz-Peaked Spectrum (GPS), through $10^{2}$ --
$10^{4}$ pc -- Compact Steep Spectrum (CSS), $10^{4}$ -- $10^{6}$ pc --
normal-size sources, up to greater than 1 Mpc -- Giant Radio Galaxies (GRG).
The three largest GRGs, recognized up to now, are shown in Fig.~1.

\begin{figure}[h]
\plotfiddle{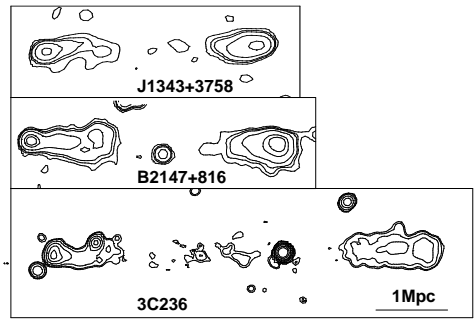}{40mm}{0}{50}{50}{-100}{0}
\caption{Three largest RGs in the Universe. A bar located in the bottom-right
corner indicates the scale of 1~Mpc.}
\end{figure}

Although giant-size radio sources are very rare among RGs, from many years 
they have been of a special interest for several reasons. Their very large 
angular size on the sky give an excellent opportunity for the study of radio
source physics. They are also very useful to study the density and evolution
of the intergalactic and intracluster environment. One of the key issues of
the current research is attempt to trace an individual evolution of RGs. Is 
there a
single evolutionary scheme governing the linear extent of radio sources, or
do small and large sources evolve in a different way? To answer this question,
in a number of papers, both theoretical and observational, attempts were
undertaken to recognize factors which may differentiate giants from 
normal-size sources. It seems that there is no a single factor responsible for 
the size
of classical radio sources; the large size of GRGs probably results from a
combination of different factors like: age of a source, jet power, density of
environment, etc. Still very limited number of well studied GRGs is a reason 
of that uncertainty. Therefore the phenomenon of GRG is still open for a
further research.

\section{Search for New Giants}
During the IAU Symposium No.~199 (December 1999) Machalski \& Jamrozy (2002)
presented an evidence
that only a very small fraction of expected faint GRGs of Fanaroff-Riley (1974)
type II (FRII) was detected at that time. In order to find those
missed giant sources we inspected the radio maps available from the large radio
surveys: NVSS (Condon et al., 1998) and the first part of FIRST (Becker et al.,
1995). The maps of these surveys, made with two different angular resolution
(45$^{\prime\prime}$ and 5$^{\prime\prime}$, respectively) at the same observing 
frequency of 1.4 GHz, allowed
(i) an effective removal of confusing sources, (ii) a reliable determination of
morphological type of the giant candidate, and (iii) a detection of the compact
radio core necessary for the proper identification of the source with its
parent optical object. As the result we selected a sample of 36 GRG candidates
(cf. Machalski et al., 2001). In order to identify their host galaxy, to
determine its distance and other physical properties, we have carried out
several radio and optical observations of the sample sources. In particular,
we already made optical spectroscopy and got redshift for 17 out of 36 galaxies
(spectroscopic redshifts of 5 sample galaxies were available prior our research).
Out of 22 galaxies, 19 host giant radio sources. In the meantime, similar efforts have
been undertaken by Schoenmakers et al. (2001) and Lara et al. (2001). Owing to
the above studies, the statistics of giant radio galaxies is enlarged. The
numbers of FRII-type GRGs, expected from our population analysis (Machalski \&
Jamrozy 2002), are recalled in Table~1 and compared with the observed numbers.
The observed numbers denoted by an asterisk refer to the data available in 
1999, while other are from the beginning of the year 2003.

\begin{table}
\caption{Statistics of FRII-type giant radio galaxies}
\begin{tabular}{lccc}
\tableline
        & $\rm 50\leq S_{1.4~GHz}<500$ mJy & $\rm 0.5\leq S_{1.4~GHz}<2$ Jy & $\rm S_{1.4~GHz}\geq 2$ Jy\\
\tableline
observed     & 64/\,11$^{*}$    & 31/\,26$^{*}$     & 11/\,9$^{*}$          \\
expected     & 350              & 45.7              & 8.8                   \\ 
obs/expected & 18\%/\,3\%$^{*}$ & 68\%/\,57\%$^{*}$ & 122\%/\,100\%$^{*}$   \\
\tableline
\tableline
\end{tabular}
\end{table}

\begin{figure}
\plottwo{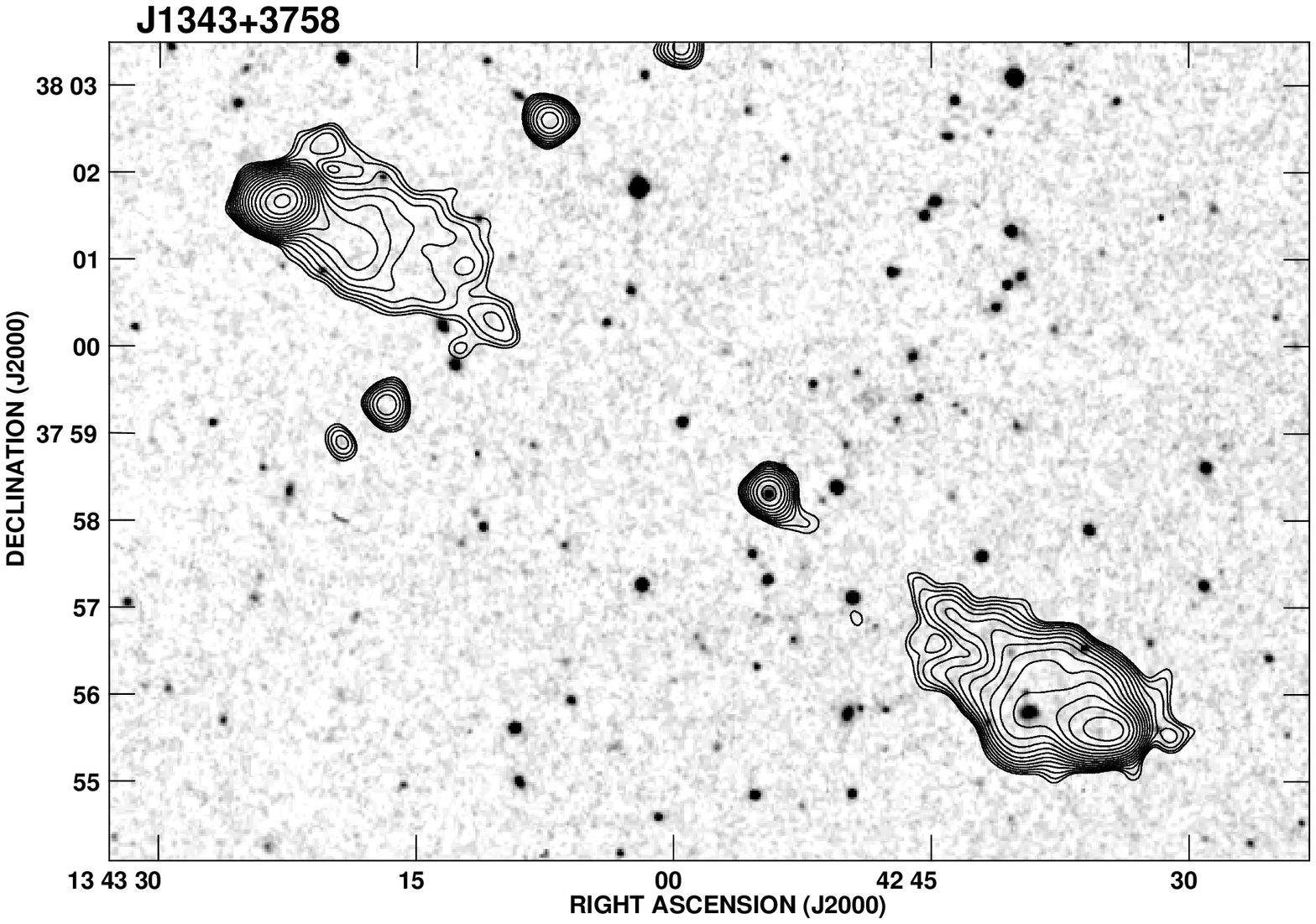}{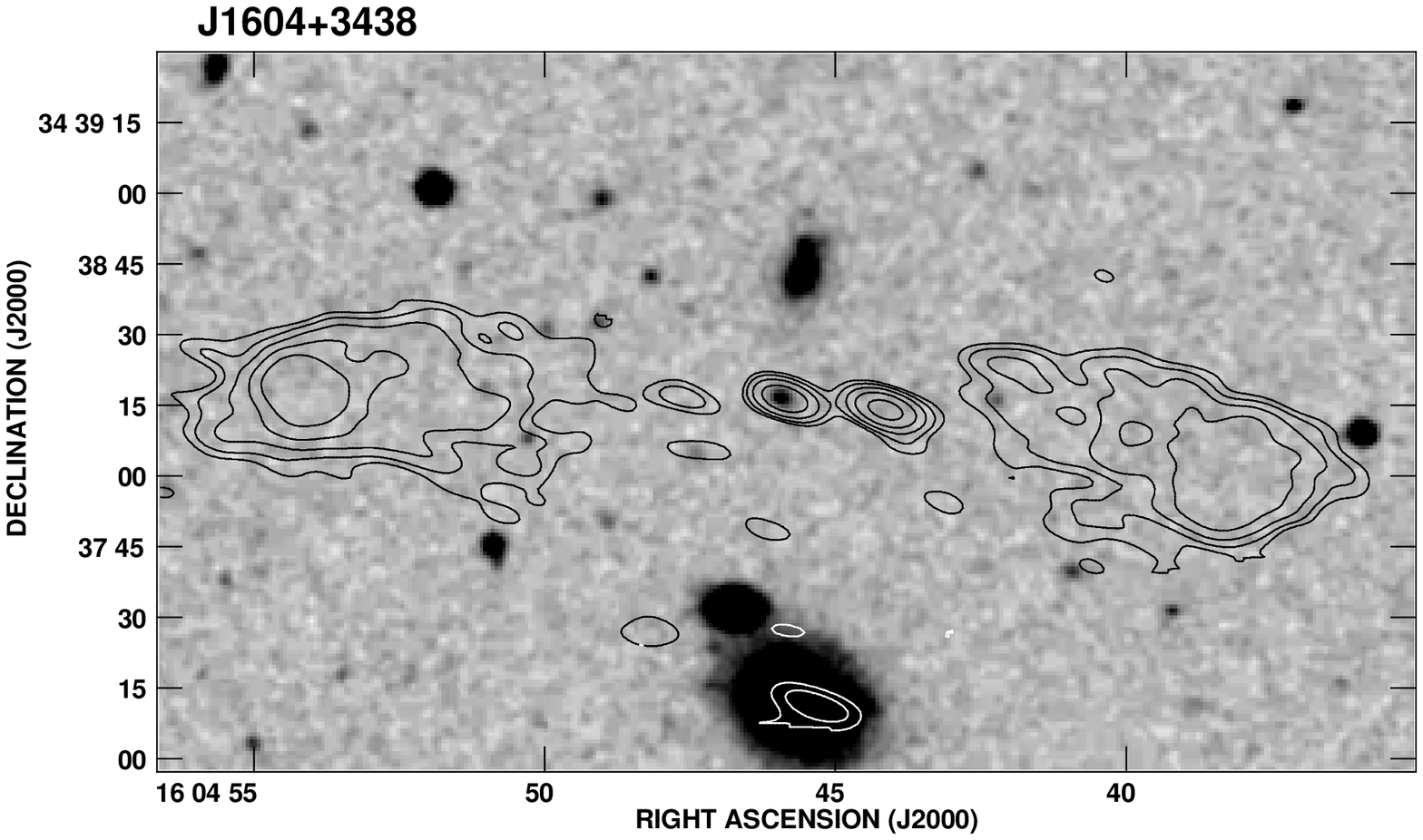}
\caption{5~GHz radio contour maps of two giant radio galaxies 
overlayed on the color optical DSS image:
J1343+3758 (left panel) and J1604+3438 (right panel).}
\end{figure}

Two examples of GRGs
from our sample are shown in Fig.~2. J1343+3758 with the linear size of
3.14 Mpc has appeared to be the third largest source in the universe (Machalski
\& Jamrozy 2000), while J1604+3438 represents a very rare type of AGN -- a
so-called double-double RG (cf. Schoenmakers et al. 2000) which shows two pairs
of lobes likely originating from an old and a new cycle of activity. 
Low-resolution optical spectra of host galaxies of these two giant radio sources 
are shown in Fig.~3.
Some of the
above data are used to constrain the existing analytical models of a dynamical
evolution of FRII-type radio sources (Machalski et al. 2003; Chy\.{z}y et al.
2003).

\begin{figure}
\plottwo{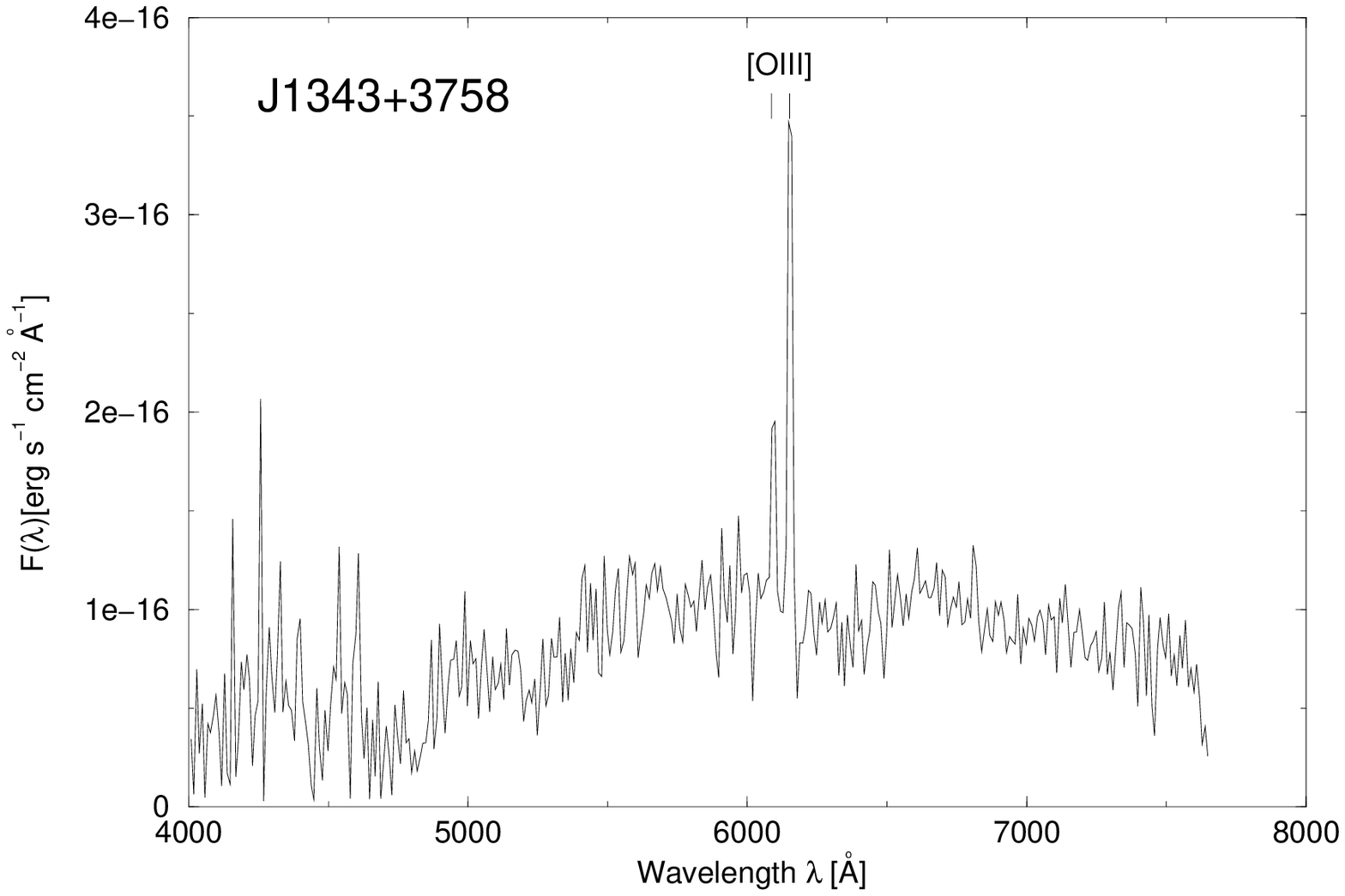}{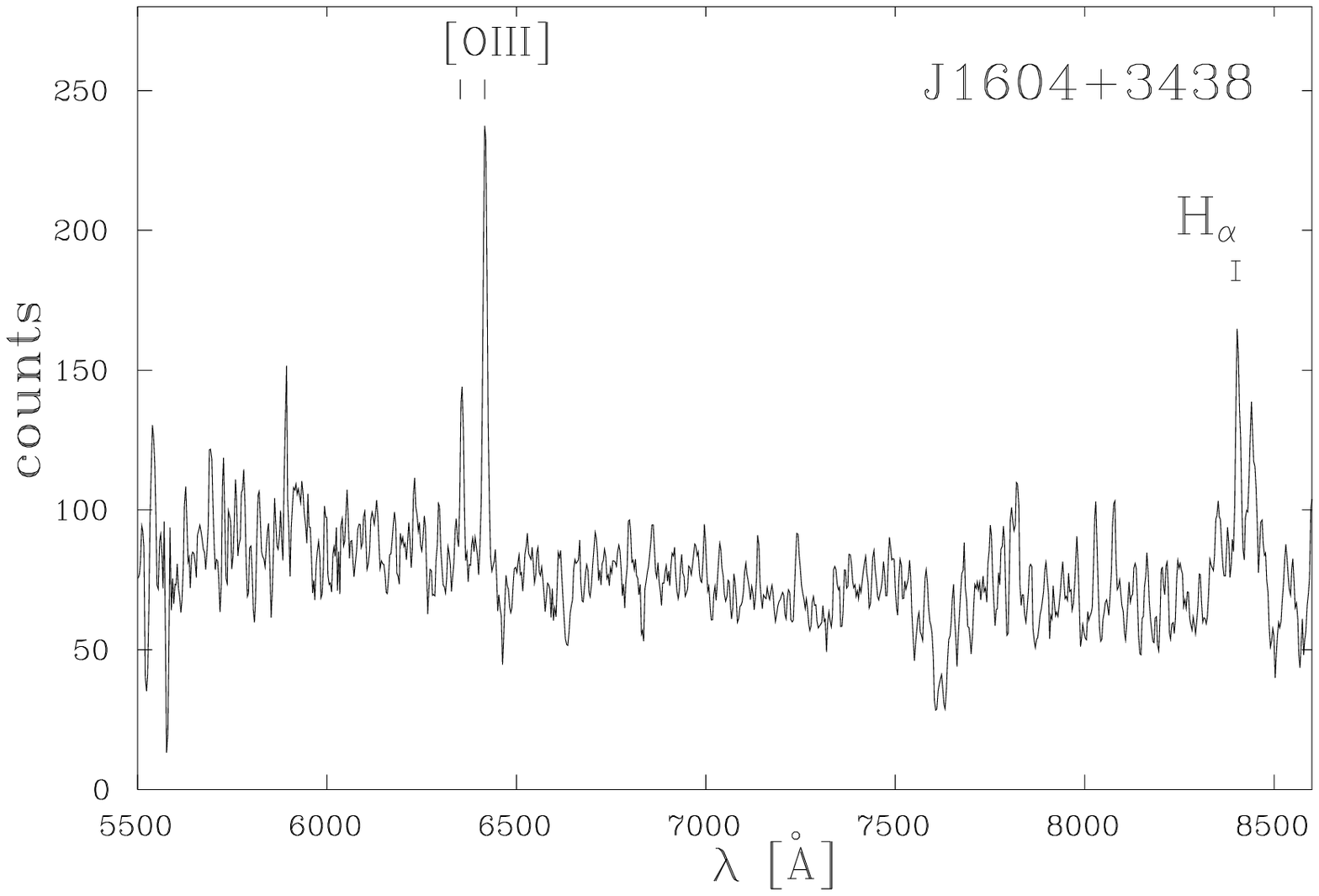}
\caption{Optical spectra of two galaxies identified with giant radio sources:
J1343+3758 (left panel) and  J1604+3438 (right panel).}
\end{figure}

\section{Why Southern Hemisphere}
Our investigations of the new giant radio sources are in progress. However, we
would like to extend them on GRGs on the southern sky. There are several scientific
reasons for such a project, and the main of them are:

\begin{itemize}
\item {All of the recent systematic search for new giants (Lara et al. 2001,
Machalski et al. 2001, Schoenmakers et al. 2001) were performed on the northern
sky. Furthermore, only about 17\% of the presently known GRGs have negative
declinations, and gross of them are high flux density ($S>$0.5 Jy) nearby
objects. Therefore, one can expect a large number of undetected yet GRGs on the
southern hemisphere very useful for a further enlargement of their still
unsatisfactory statistics.}
\item {The  development of astronomical high-technology facilities i.e. the
existing and planned large optical telescopes on the South (VLT, SALT) is very
rapid. Therefore, it should be easy to attain the redshift of new GRG hosting
galaxies which is the crucial observational parameter for determination of all
physical parameters of the radio sources like their distance, projected linear
size, volume of their lobes or cocoon, luminosity, etc. The above needs
low-resolution spectroscopic observations of usually faint optical counterparts
(which in many cases have very low apparent magnitudes $R\sim21^{m}-23^{m}$) in
a reasonably short time.}
\item {There is a high probability that the planned powerful radio interferometer --
the Square Kilometer Area (SKA)  will be located in the South (there are two
southern site candidates for its possible location, i.e. Australia or South
Africa). It would give the opportunity to detect and study low 
surface-brightness GRGs with a very high angular resolution and sensitivity in
the future. This, in turn, would allow to recognize a very last stage of the
dynamical evolution of classical double radio galaxies and learn a typical
lifetime and end of that physical process.}
\end{itemize}

\end{document}